\documentclass[11pt]{article}

\usepackage[final]{acl}
\usepackage{times}
\usepackage{latexsym}

\usepackage[T1]{fontenc}
\usepackage[utf8]{inputenc}
\usepackage{microtype}
\usepackage{inconsolata}
\usepackage{courier}
\usepackage{natbib}
\usepackage{caption}
\usepackage{amsmath}
\usepackage{amssymb}
\usepackage{booktabs}

\usepackage{pifont}
\usepackage{multirow}

\usepackage{graphicx}
\usepackage[nameinlink]{cleveref}
\crefname{section}{Sec.}{Secs.}
\Crefname{section}{Sec.}{Secs.}
\crefname{subsection}{Sec.}{Secs.}
\Crefname{subsection}{Sec.}{Secs.}
\crefname{subsubsection}{Sec.}{Secs.}
\Crefname{subsubsection}{Sec.}{Secs.}
\crefname{table}{Tab.}{Tabs.}
\Crefname{table}{Tab.}{Tabs.}
\crefname{equation}{Eq.}{Eqs.}
\Crefname{equation}{Eq.}{Eqs.}
\crefname{figure}{Fig.}{Figs.}
\Crefname{figure}{Fig.}{Figs.}
\crefname{appendix}{App.}{App.}
\Crefname{appendix}{App.}{App.}

\title{{PUMA}: Post-Hoc Sparsification of Universal \\Multimodal Embeddings for Efficient Retrieval}

\author{
{\bfseries
Matteo Attimonelli$^{1,2}$,
Alessandro De Bellis$^{1}$,
Franco Maria Nardini$^{3}$,
Claudio Pomo$^{1}$} \\
{\bfseries
Cosimo Rulli$^{3}$,
Rossano Venturini$^{4}$,
Tommaso Di Noia$^{1}$} \\
\\
$^{1}$Politecnico di Bari, Italy \quad
$^{2}$Sapienza University of Rome, Italy \\
$^{3}$ISTI--CNR, Pisa, Italy \quad
$^{4}$University of Pisa, Italy \\
{\small
\textbf{Correspondence:} \texttt{matteo.attimonelli@poliba.it}}
}

\begin{document}
\maketitle

\begin{abstract}
Universal multimodal embedders enable retrieval across text, image, and combined queries, but their dense representations incur high memory and inference costs. Post-hoc sparsification could reduce these costs but remains underexplored for multimodal retrieval.
We introduce \textbf{PUMA}, a sparse autoencoder recipe that maps universal multimodal embeddings to compact sparse codes without retraining the backbone: a pretraining stage preserves dense dot-product geometry, after which the sparse encoder is fine-tuned for retrieval.
We evaluate on five benchmarks covering text-to-image and composed image retrieval.
On Qwen3-VL-Embedding-2B, PUMA is statistically indistinguishable from or improves over dense retrieval on four of five datasets. We further identify two failure modes of post-hoc sparsification: insufficient pre-TopK support and retrieval-misaligned active support.
PUMA reduces vector storage by $8$--$16\times$ (FP32) and is up to $25\times$ faster than exact dense scoring on larger candidate pools, enabling efficient multimodal retrieval.
Code is provided at our \href{https://github.com/sisinflab/PUMA}{ \texttt{GitHub Repository}}.
\end{abstract}

\section{Introduction}
Universal multimodal embedders offer a practical unified interface for retrieval across text, image, and mixed-modality queries \citep{mmembed,DBLP:conf/cvpr/ZhangZXLDLXZLZ25}. A single backbone maps queries and candidates of any modality into one shared space, without separate retrieval heads or fusion modules. This generality comes at a cost: embeddings are dense, high-dimensional, and expensive to store and score over large candidate sets. Exact dense scoring over $n$ candidates with $d$-dimensional embeddings costs $O(nd)$ per query, and approximate dense retrieval still requires specialized vector indexes over large floating-point representations. Sparse retrieval offers a complementary alternative, storing vectors as small sets of components and scoring through feature overlap.

Prior works studied sparse text retrieval \citep{splade,csr,splare,csr_v2,milco}, sparsity during multimodal contrastive pretraining \citep{sparseclip}, and post-hoc multimodal sparse autoencoders for representation analysis \citep{mgsae}.
What remains underexplored is post-hoc sparsification for retrieval serving: can a frozen universal multimodal embedder be converted into a sparse retriever without retraining the backbone, while preserving quality across diverse tasks?

We propose \textbf{PUMA} (\textbf{P}ost-hoc \textbf{U}niversal \textbf{M}ultimodal sparsific\textbf{A}tion), a post-hoc sparsification pipeline for frozen universal multimodal embedders. PUMA converts cached dense embeddings into sparse retrieval codes using a TopK sparse autoencoder. PUMA is first trained to preserve the dense dot-product geometry and is then fine-tuned with the sparse dot-product used at retrieval time. The resulting representation has exactly $k$ active features per vector and can be scored as a sparse dot product.
To the best of our knowledge, PUMA is the first work to study post-hoc sparsification of frozen universal multimodal embedders for retrieval serving.

We evaluate PUMA on five M-BEIR datasets \citep{mbeir}, with candidate pools ranging from roughly $5$K to $542$K images. The tasks span two regimes: \emph{text-to-image retrieval}, where text queries retrieve image candidates, and \emph{composed image retrieval}, where a reference image and textual modification are fused into a single query representation. The latter is especially challenging as sparsification must preserve the modification semantics of this fused image--text query while aligning it with target images.

On Qwen3-VL-Embedding-2B~\citep{qwen3vl}, PUMA significantly improves over dense retrieval on CIRR and Fashion200K, is statistically indistinguishable on FashionIQ and MSCOCO, and trails dense on VisualNews. On CIRR, PUMA reaches $0.533$ Hit@5 and $0.436$ nDCG@10, compared with $0.522$ and $0.424$ for dense retrieval.

These gains are not explained by dimensionality reduction or domain-specific finetuning alone. In fact, PUMA outperforms dense TopK pruning, matched-memory PCA, a trained dense autoencoder, and encoder-only sparse training.
Compared to a dense retriever, PUMA uses $8$--$16\times$ less per-vector storage and makes exact sparse search up to $25\times$ faster on larger candidate pools.

Our analysis identifies two ways in which post-hoc sparsification can fail. In some cases, the sparse encoder produces insufficient active features before TopK selection, so increasing $k$ cannot help. In other cases, enough features are active, but they do not preserve the original dense ranking behavior. These failure modes suggest that future sparse multimodal retrievers must control both feature availability and retrieval alignment.

\noindent We make the following contributions:
\begin{enumerate}\setlength\itemsep{0pt}
\item We formulate post-hoc sparsification of frozen universal multimodal embedders as a retrieval problem, motivated by the storage and scoring costs of dense multimodal representations.
\item We propose \textbf{PUMA}, a retrieval-aware TopK sparse-autoencoder recipe that converts dense multimodal embeddings into compact sparse codes without retraining the backbone.
\item We evaluate PUMA on five datasets spanning text-to-image and composed image retrieval, showing competitive performance with dense baselines at substantially lower storage cost.
\item We compare against dense TopK pruning, matched-memory PCA, a dense autoencoder, and encoder-only sparse training, and analyze search time, statistical reliability, scaling, transfer, and $k$-selection failure modes.
\end{enumerate}

Our results show that post-hoc sparsification offers a practical path to efficient multimodal retrieval, preserving the generality of universal embedders at a fraction of the storage and search cost.

\section{Related Work}
\label{sec:related}

\paragraph{Universal multimodal retrieval.}
Universal multimodal retrieval aims to use a single embedding model across heterogeneous query--candidate pairs, including cross-modal and mixed-modality queries. M-BEIR~\citep{mbeir} provides a unified benchmark across tasks and modalities, and MMEB~\citep{DBLP:conf/iclr/JiangMYYZC25} broadens evaluation to new instruction-conditioned tasks. Recent dense embedders such as VLM2Vec~\citep{DBLP:conf/iclr/JiangMYYZC25}, MM-Embed~\citep{mmembed}, GME~\citep{DBLP:conf/cvpr/ZhangZXLDLXZLZ25}, Qwen3-VL-Embedding~\citep{qwen3vl}, and RZenEmbed~\citep{DBLP:journals/corr/abs-2510-27350} adapt multimodal LLMs or vision-language models into general-purpose embedding backbones; SERVAL~\citep{serval} explores a complementary generate-and-encode route for visual document retrieval.
PUMA is orthogonal to these works: rather than training a new model, we study whether pretrained dense representations can be sparsified post-hoc. We focus on universal multimodal embedders as they produce shared representations for mixed-modality inputs without modality-specific encoders or fusion modules.

\begin{figure*}[t]
    \centering
    \includegraphics[width=\linewidth]{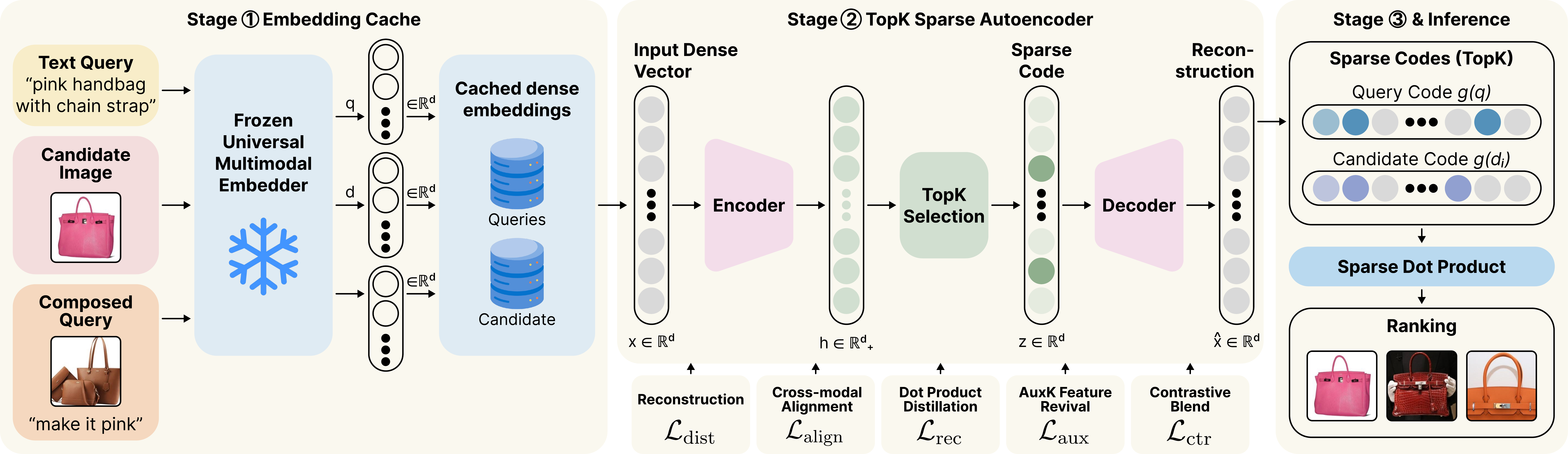}
    \caption{PUMA: (i) embeddings are cached using frozen embedders; (ii) a top-$K$ SAE is trained to produce sparse representations; (iii) these representations are further tuned to optimize retrieval, and ranking is performed.
    }
\label{fig:puma}
\end{figure*}

\paragraph{Sparse retrieval and embedding compression.}
Dense retrieval efficiency can be improved through quantization~\citep{pq}, approximate search~\citep{scann}, or dense dimensionality reduction~\citep{DBLP:conf/recsys/AttimonelliBPJS25}, but these approaches keep dense representations or change the search procedure. Sparse retrieval instead changes the representation structure, storing feature--weight pairs that can reuse lexical-search infrastructure.
This paradigm has been studied extensively for text, both in lexical spaces~\citep{splade,splade_v2} and in latent sparse codes, including CSR~\citep{csr,csr_v2}, MILCO~\citep{milco}, and SPLARE~\citep{splare}.
Sparsification has also been explored in recommendation, including sparse contrastive supervision and sparse entity embeddings~\citep{DBLP:journals/corr/abs-2604-12990,DBLP:conf/www/VancuraSAP26}.
PUMA builds on $k$-sparse autoencoders~\citep{makhzani_ktopk} and recent SAE work in representation analysis~\citep{cunningham_sae,bricken_monosemanticity,gao_topksae}.
The closest multimodal comparisons differ in goal and intervention point: Sparse CLIP~\citep{sparseclip} optimizes sparsity during contrastive pretraining, whereas PUMA sparsifies a pretrained embedder post-hoc; MGSAE~\citep{mgsae} studies post-hoc SAEs for interpretability, whereas PUMA targets retrieval.
PUMA adapts sparse autoencoding to frozen universal multimodal embedders by preserving dense retrieval geometry and fine-tuning the sparse encoder with a retrieval objective. Importantly, PUMA uses a shared multimodal representation rather than modality-specific ones.

\section{PUMA}
\label{sec:method}

\Cref{fig:puma} gives an overview of \textbf{PUMA}, a post-hoc sparsification approach for frozen multimodal embedders.
PUMA converts dense embeddings from a frozen multimodal backbone into sparse retrieval codes. It learns an overcomplete dictionary of $D$ latent features, with $D \gg d$: a TopK encoder maps a dense embedding $\mathbf{x}\in\mathbb{R}^d$ to a sparse code $\mathbf{z}\in\mathbb{R}^D$ with exactly $k$ nonzero entries ($||z||_0 = k)$, and a linear decoder reconstructs $\hat{\mathbf{x}}$. Training has three stages. Stage~\ding{172} caches dense embeddings from the frozen multimodal encoder. Stage~\ding{173} pretrains the sparse autoencoder with reconstruction, retrieval-aware distillation, cross-modal alignment, feature revival for inactive dictionary atoms, and progressive $k$-annealing. Stage~\ding{174} fine-tunes the encoder with a contrastive loss on sparse codes, using the same sparse dot product used at retrieval time.

\subsection{Problem Setting}
Let $f(\cdot)$ be a frozen multimodal embedder that maps queries and candidates into a shared dense space. A query may be text only, image only, or composed from a reference image and a textual modification. We write $\mathbf{q}=f(\text{query})$ and $\mathbf{d}=f(\text{candidate})$, and dense retrieval ranks candidates by $\mathbf{q}^{\top}\mathbf{d}$. We learn a sparse encoder $g(\cdot)$ that maps any dense embedding to
\begin{equation}
\mathbf{z} = g(\mathbf{x}) \in \mathbb{R}^{D}, \qquad \|\mathbf{z}\|_{0}=k,
\end{equation}
where $D$ is the size of the learned sparse dictionary, $D \gg d$, and $k \ll D$ is the number of active dictionary atoms per vector.
Sparse retrieval ranks candidates by
\begin{equation}
s_{\mathrm{sparse}}(\mathbf{q},\mathbf{d}) = g(\mathbf{q})^{\top}g(\mathbf{d}).
\end{equation}
The goal is to preserve dense retrieval rankings under a compact sparse representation.

\subsection{TopK Sparse Autoencoder (SAE)}
The sparse autoencoder computes, for each query and candidate embedding:
\begin{align}
\mathbf{h} &= \mathrm{ReLU}(W_{\mathrm{enc}}\mathbf{x} + \mathbf{b}_{\mathrm{enc}}),\\
\mathbf{z} &= g(\mathbf{x}) = \mathrm{TopK}(\mathbf{h}, k),\\
\hat{\mathbf{x}} &= W_{\mathrm{dec}}\mathbf{z} + \mathbf{b}_{\mathrm{dec}}.
\end{align}
Decoder ($W_{\text{dec}}$) columns are renormalized to unit norm after each optimization step to prevent scale drift. For our experiments, we set $D{=}16d$.

\subsection{Stage \ding{172}: Embedding Cache}
We run the frozen backbone $f$ on the train, validation, and candidate-pool partitions of each task and store the resulting dense embeddings. All subsequent sparse training reads from this cache. This makes the training cost independent of the backbone and ensures that dense and sparse variants use identical underlying embeddings.

\subsection{Stage \ding{173}: Retrieval-Aware Sparse Pretraining}
\label{sec:stage2}

A reconstruction objective alone is not directly aligned with ranking. Moreover, post-hoc sparse autoencoders over multimodal embeddings can learn modality-specific dictionaries, where different atoms fire for images and text, weakening cross-modal scoring \citep{mgsae}. Stage~2 therefore trains on cached in-domain query--target pairs with the objective:
\begin{equation}
\label{eq:stage2}
\mathcal{L}_{\mathrm{2}} =
\mathcal{L}_{\mathrm{rec}}
+ \lambda_{\mathrm{a}}\mathcal{L}_{\mathrm{align}}
+ \lambda_{\mathrm{d}}\mathcal{L}_{\mathrm{dist}}
+ \lambda_{\mathrm{x}}\mathcal{L}_{\mathrm{aux}}
+ \lambda_{\mathrm{c}}\mathcal{L}_{\mathrm{ctr}}.
\end{equation}
\paragraph{Reconstruction.}
We reconstruct both query and candidate embeddings with cosine reconstruction:
\begin{equation}
\mathcal{L}_{\mathrm{rec}} = 1 - \cos(\mathbf{x},\hat{\mathbf{x}}).
\end{equation}

\paragraph{Cross-modal alignment.}
To make dictionary usage more balanced across modalities, we encourage paired cross-modal examples to activate overlapping atoms instead of assigning image and text embeddings to disjoint subsets of the $D$-dimensional sparse space. For an image/text pair
$(\mathbf{x}_{\mathrm{img}},\mathbf{x}_{\mathrm{txt}})$ with sparse codes $\mathbf{z}_{\mathrm{img}}$ and $\mathbf{z}_{\mathrm{txt}}$, we use
\begin{equation}
\normalsize
\mathcal{L}_{\mathrm{align}}\mkern-5mu=\mkern-5mu
\|\mathbf{m}_{\mathrm{img}}-\mathbf{m}_{\mathrm{txt}}\|_1+
\frac{1}{2}\sum_{j\in\mathcal{A}}
\big||z_{\mathrm{img},j}|-|z_{\mathrm{txt},j}|\big|.
\end{equation}
where $\mathbf{m}_{\mathrm{img}}$ and $\mathbf{m}_{\mathrm{txt}}$ denote the TopK support masks, and $\mathcal{A}=\{j:z_{\mathrm{img},j}\neq0 \wedge z_{\mathrm{txt},j}\neq0\}$ is the co-active support. The first term encourages shared support; the second aligns magnitudes on atoms that fire for both modalities. The support masks come from the hard TopK forward pass; gradients for the support-alignment term are passed to the selected pre-TopK activations using the same straight-through treatment as the TopK operation~\citep{mgsae}. For retrieval pairs, we apply this alignment to cross-modal query--target pairs: in text-to-image retrieval, this aligns the text-query code with the positive image-candidate code; in composed image retrieval, it aligns the composed image--text query code with the positive target-image code.

\paragraph{Dot-product distillation.}
We distill the dense retriever's scoring geometry into sparse space, matching dense and sparse scores for each query--candidate pair:
\begin{equation}
\mathcal{L}_{\mathrm{dist}} =
\left(g(\mathbf{q})^{\top}g(\mathbf{d}) - \mathbf{q}^{\top}\mathbf{d}\right)^2.
\end{equation}

\paragraph{Auxiliary feature revival (AuxK).}
A TopK encoder gives gradient only to the $k$ selected atoms, so the remaining $D-k$ tend to die. Following \citet{gao_topksae,csr_v2}, we revive them by asking the dead atoms to reconstruct the residual the main code missed. Let $\mathcal{G}\subset\{1,\ldots,D\}$ be the set of atoms inactive for more than $T$ steps, and form an auxiliary code on the dead support $\mathbf{z}_{\mathrm{aux}} = \mathrm{TopK}_{k_{\mathrm{aux}}}(\mathrm{softplus}(\mathbf{h}|_{\mathcal{G}}))$:
\begin{equation}
\mathcal{L}_{\mathrm{aux}} =
\big\| W_{\mathrm{dec}}\mathbf{z}_{\mathrm{aux}} - \mathrm{sg}\!\left[\mathbf{x} - \hat{\mathbf{x}}\right] \big\|_2^2,
\end{equation}
where $\mathrm{sg}[\cdot]$ is the stop-gradient operator and ensures that only dead atoms are updated by this term.

\paragraph{Stage~\ding{173} contrastive blend.}
We include a small InfoNCE term on sparse codes during pretraining,
\begin{equation}
\mathcal{L}_{\mathrm{ctr}} =
-\log
\frac{
\exp(g(\mathbf{q}_i)^{\top}g(\mathbf{d}_i)/\tau)
}{
\sum_j \exp(g(\mathbf{q}_i)^{\top}g(\mathbf{d}_j)/\tau)
},
\end{equation}
which begins aligning the sparse representation with ranking before retrieval fine-tuning.

\paragraph{Progressive $k$-annealing.}
We start from a larger support $k_{\mathrm{init}}$ and linearly anneal to the target support $k_{\mathrm{final}}$. This gives the dictionary a denser training signal early and ends training at the sparsity used for retrieval~\citep{csr_v2}.

\subsection{Stage \ding{174}: Contrastive Retrieval Fine-Tuning}
Stage~\ding{174} optimizes the sparse encoder for retrieval with the same dot product used at inference. We use InfoNCE with in-batch negatives and, when available, explicit hard negatives, while retaining reconstruction and distillation as regularizers:
\begin{equation}\label{eq:stage3}
\mathcal{L}_{\mathrm{3}} =
\mathcal{L}_{\mathrm{ctr}}
+ \alpha\mathcal{L}_{\mathrm{rec}}
+ \beta\mathcal{L}_{\mathrm{dist}},
\end{equation}
with $\alpha{=}0.1$ and $\beta{=}0.05$ in the main experiments. This stage shifts the model from reconstruction-oriented toward retrieval-oriented sparse scoring.

\section{Experimental Setup}
\label{sec:setup}

We evaluate on five M-BEIR datasets \citep{mbeir}. Two are composed image retrieval tasks with image-plus-text queries and image candidates: \textbf{CIRR} (26{,}116 train queries, 4{,}170 test queries, 21{,}551-image candidate pool) \citep{cirr} and \textbf{FashionIQ} (6{,}003 test queries, 74{,}381-image candidate pool) \citep{fashioniq}. Three are text-to-image retrieval tasks: \textbf{VisualNews} (19{,}996 queries, 542{,}246 candidates), \textbf{Fashion200K} (1{,}746 queries, 201{,}824 candidates), and \textbf{MSCOCO} (24{,}819 queries, 5{,}000 candidates). We focus on these two regimes because they capture complementary aspects of multimodal retrieval: text-to-image retrieval is the standard cross-modal setting, while composed image retrieval tests whether a model can represent a genuinely multimodal query, combining visual context from a reference image with a textual modification. Together, the five tasks span candidate pools from roughly $5$K to $542$K images.

\paragraph{Metrics.}
We report Hit@5, Recall@10, and nDCG@10. Hit@5 measures whether at least one relevant candidate appears in the top 5 and is the standard headline metric for CIRR and FashionIQ. Recall@10 measures coverage at a deeper cutoff, while nDCG@10 weights both relevance and rank position. We use nDCG@10 for validation.

\paragraph{Backbones.}
We focus on universal multimodal embedding models because they directly output shared embeddings for text, images, and image--text inputs, without requiring late fusion of modality-specific representations. They also represent the current strong-performing class of multimodal retrievers. Our main experiments use Qwen3-VL-Embedding-2B (Qwen-2B)~\citep{qwen3vl}, which produces $d{=}2048$ dimensional embeddings. We also evaluate Qwen3-VL-Embedding-8B (Qwen-8B, $d{=}4096$)~\citep{qwen3vl} for within-family scaling and RZenEmbed~\citep{DBLP:journals/corr/abs-2510-27350} ($d{=}3584$) for cross-family transfer.
Dense baselines and all sparse variants use identical prompts, cached embeddings, and candidate pools, to isolate sparsification effects.

\paragraph{Baselines.}
We compare PUMA against four baselines. \emph{Full dense} retrieval ranks candidates with the original frozen embeddings. \emph{Raw TopK dense} keeps the largest-magnitude dense coordinates with signed weights, representing a na\"ive sparsification baseline. \emph{Matched-memory PCA} projects dense embeddings to dimension $2k$ using PCA fit on the candidate pool, approximately matching the storage of a sparse code with support $k$ under FP32 $(\text{id},\text{weight})$ accounting. \emph{Trained dense AE} is a linear autoencoder with bottleneck dimension $2k$, trained on the same data with the same retrieval losses, optimization budget, and validation protocol as PUMA where applicable; sparsity-specific components such as AuxK, TopK support annealing, and support-mask alignment are omitted or replaced by dense analogs. \emph{Encoder-only TopK} keeps the TopK sparse encoder and retrieval supervision but removes the decoder and reconstruction path, testing whether the autoencoding structure contributes beyond sparse retrieval supervision alone.

\paragraph{Implementation details.}
All sparse models are trained on cached dense embeddings with the backbone frozen, at expansion ratio $16\times$ (so $D{=}32{,}768$ for Qwen-2B and analogously for larger backbones). Dense embeddings are extracted with vLLM when supported by the backbone, and sparse training/evaluation is implemented in PyTorch. All experiments are run on a machine equipped with a single NVIDIA H100 NVL GPU. We use linear warmup with cosine learning-rate decay, decoder column normalization after each step, and progressive $k$-annealing to the target support. Checkpoints are selected by held-out sparse nDCG@10; PUMA operating points are chosen on validation and evaluated once on test. Storage and search-time accounting, along with full hyperparameters and hardware details, are reported in \cref{sec:results-efficiency} and \cref{app:training-details}.

\begin{table*}[t]
\caption{Main results across three backbones (Qwen-2B, Qwen-8B, RZenEmbed). \textbf{Bold} = better of Dense/PUMA per (data, backbone, metric). PUMA $k$ for Qwen-2B and RZenEmbed: 160 (CIRR/FashIQ/VisN), 128 (F200K), 144 (MSCOCO); for Qwen-8B: 144 (CIRR), 128 (elsewhere).}
\label{tab:main-results}
\centering
\small
\setlength{\tabcolsep}{8pt}
\resizebox{\textwidth}{!}{
\begin{tabular}{@{}ll ccc ccc ccc@{}}
\toprule
& & \multicolumn{3}{c}{\textbf{Qwen-2B}} & \multicolumn{3}{c}{\textbf{Qwen-8B}} & \multicolumn{3}{c}{\textbf{RZenEmbed}} \\
\cmidrule(lr){3-5}\cmidrule(lr){6-8}\cmidrule(lr){9-11}
Data & Method & H@5 & R@10 & N@10 & H@5 & R@10 & N@10 & H@5 & R@10 & N@10 \\
\midrule
\multirow{2}{*}{CIRR}        & Dense & .5221 & .6281 & .4235 & \textbf{.5570} & \textbf{.6577} & \textbf{.4961} & \textbf{.5930} & \textbf{.6914} & \textbf{.4860} \\
                             & PUMA  & \textbf{.5329} & \textbf{.6449} & \textbf{.4356} & .5392 & .6438 & .4835 & .5664 & .6743 & .4713 \\
\midrule
\multirow{2}{*}{FashionIQ}   & Dense & .2029 & .2680 & \textbf{.1706} & .2321 & .3027 & \textbf{.1942} & .2342 & .3061 & .1952 \\
                             & PUMA  & \textbf{.2062} & \textbf{.2697} & .1681 & \textbf{.2348} & \textbf{.3055} & .1922 & \textbf{.2424} & \textbf{.3144} & \textbf{.1994} \\
\midrule
\multirow{2}{*}{VisualNews}  & Dense & \textbf{.2666} & \textbf{.3318} & \textbf{.2256} & .2720 & .3396 & .2296 & \textbf{.4587} & \textbf{.5410} & \textbf{.3933} \\
                             & PUMA  & .2590 & .3243 & .2168 & \textbf{.3052} & \textbf{.3820} & \textbf{.2561} & .4215 & .5069 & .3588 \\
\midrule
\multirow{2}{*}{Fashion200K} & Dense & .1198 & .0879 & .0792 & .1646 & .1215 & .1048 & .2054 & .1521 & .1339 \\
                             & PUMA  & \textbf{.1763} & \textbf{.1355} & \textbf{.1144} & \textbf{.2071} & \textbf{.1595} & \textbf{.1318} & \textbf{.2478} & \textbf{.1813} & \textbf{.1579} \\
\midrule
\multirow{2}{*}{MSCOCO}      & Dense & .8049 & .8721 & .7181 & .8184 & .8835 & \textbf{.7349} & \textbf{.8447} & \textbf{.9012} & \textbf{.7630} \\
                             & PUMA  & \textbf{.8098} & \textbf{.8805} & \textbf{.7191} & \textbf{.8220} & \textbf{.8871} & .7297 & .8270 & .8884 & .7384 \\
\bottomrule
\end{tabular}}
\end{table*}

\section{Results}
\label{sec:results}

We first evaluate PUMA on Qwen3-VL-Embedding-2B across all five M-BEIR tasks, then test whether the result transfers to Qwen3-VL-Embedding-8B and RZenEmbed. We next compare against simpler dense and sparse compression baselines, analyze storage and search time, and ablate the main training components. Unless otherwise stated, significance is measured with 95\% bootstrap intervals and 1000-trial paired approximate-randomization tests on nDCG@10.

\subsection{Main Results on Qwen-2B}
\cref{tab:main-results} reports results on the five M-BEIR tasks using Qwen3-VL-Embedding-2B. PUMA is competitive with dense retrieval across the benchmark family: it significantly improves over dense retrieval on CIRR and Fashion200K, is statistically indistinguishable on FashionIQ and MSCOCO, and trails dense only on VisualNews.
On CIRR, PUMA improves nDCG@10 from $0.4235$ to $0.4356$ ($p\!\le\!0.001$), while also improving Hit@5 and Recall@10. On Fashion200K, the gain is larger, with nDCG@10 increasing from $0.0792$ to $0.1144$ ($p\!\le\!0.001$). On FashionIQ and MSCOCO, the differences are small and not statistically reliable. VisualNews is the only Qwen-2B setting where dense retrieval remains clearly ahead.
Overall, PUMA matches or exceeds dense retrieval on four of five tasks by the primary criterion of competitive or better nDCG@10, while using substantially smaller sparse representations. Representative bootstrap intervals are reported in \cref{app:bootstrap}.

\subsection{Scaling and Cross-Family Transfer}
\label{sec:results-generalization}

To test generalization, we use the same sparsification recipe on Qwen3-VL-Embedding-8B and RZenEmbed, changing only cached embeddings.

\paragraph{Within-family scaling: Qwen-8B.}
On Qwen3-VL-Embedding-8B (\cref{tab:main-results}), PUMA improves over dense retrieval on the two larger text-to-image datasets, VisualNews and Fashion200K, by about $+0.027$ nDCG@10 in both cases ($p\!\le\!0.001$). It remains close to dense on MSCOCO and FashionIQ, but falls below dense on CIRR. Thus, scaling the dense backbone strengthens PUMA on text-to-image retrieval, but does not automatically make composed retrieval easier to sparsify.

\paragraph{Cross-family transfer: RZenEmbed.}
\label{sec:results-rzen}
On RZenEmbed (\cref{tab:main-results}), PUMA exceeds dense on FashionIQ and Fashion200K, with the Fashion200K gain statistically reliable ($p\!\le\!0.001$) and FashionIQ trending positive ($p{=}0.075$). It remains below dense on CIRR, VisualNews, and MSCOCO.
This shows that transfer is real but uneven across backbones. More broadly, the RZenEmbed results suggest that the PUMA recipe is not specific to the Qwen family, but that its success depends on backbone geometry and task regime. In particular, the mixed transfer pattern is consistent with the two-way diagnosis developed below: some settings fail because support is not available before TopK, while others fail because the available support is not retrieval-aligned enough.

\paragraph{Failure diagnostics.}
The two hardest settings fail for different reasons (\cref{tab:failure-diagnostics}). On Qwen-8B CIRR, the encoder produces only 166.5 positive pre-activations on average and realizes $k{=}160$ on only 81.8\% of examples, so the requested support is not always available before TopK. On RZen VisualNews, support is available: the model produces 465.3 positive pre-activations on average and reaches $k{=}160$ on every example. Yet sparse nDCG@10 remains well below dense, indicating that the active features are not retrieval-aligned. Thus support availability is necessary but not sufficient; the active support must also preserve dense retrieval geometry. These diagnostics make the failure modes interpretable: post-hoc sparsification can fail either because the encoder does not produce enough pre-TopK support, or because the available support does not preserve retrieval rankings.
Both diagnostics can be monitored during training and point to objectives that jointly improve dictionary coverage and retrieval alignment.

\begin{table}[t]
\caption{Support diagnostics for two hard settings. Pos-pre: mean positive pre-TopK activations; Reach: fraction realizing $k$ exactly; Used: distinct active features; PUMA/Dn: PUMA/dense nDCG@10.}
\label{tab:failure-diagnostics}
\centering
\small
\setlength{\tabcolsep}{5.5pt}
\begin{tabular}{@{}lccccc@{}}
\toprule
Setting & $k$ & Pos-pre & Reach & Used & PUMA/Dn \\
\midrule
8B CIRR   & 160 & 166.5 & 0.818 & 199 & .484/.496 \\
RZen VisN & 160 & 465.3 & 1.000 & 497 & .359/.393 \\
\bottomrule
\end{tabular}
\end{table}

\subsection{Alternative Compression Techniques}
\label{sec:results-baselines}

We compare PUMA with four compression baselines designed to separate sparsity, memory budget, adaptation, and autoencoding effects, along with the original backbone without compression (Dense). Raw TopK dense coordinates test whether direct coordinate pruning is enough; matched-memory PCA tests whether a smaller dense representation suffices; the trained dense autoencoder tests whether matched-budget learned dense adaptation closes the gap;
and encoder-only TopK keeps the sparse encoder but removes the decoder, testing whether sparse supervision alone is sufficient.

\cref{tab:baselines-2b} compares PUMA on three representative Qwen-2B settings against raw TopK dense coordinates, matched-memory PCA, and an encoder-only TopK sparse model. Raw TopK is substantially weaker: on CIRR, it reaches $0.3769$ nDCG@10 versus PUMA's $0.4356$. Matched-memory PCA is stronger ($0.4153$) but remains below both dense and PUMA. The encoder-only sparse model — same TopK encoder and retrieval supervision as PUMA but without a decoder or reconstruction pathway — reaches $0.3631$ on CIRR, above raw TopK but clearly below PUMA. The pattern reproduces on FashionIQ and Fashion200K. None of the three explanations holds: coordinate sparsity alone is insufficient, memory reduction alone is insufficient, and sparse retrieval supervision without a reconstruction signal is insufficient.

\begin{table}[t]
\caption{Alternative compressors vs.\ PUMA vs.\ base uncompressed model (Dense) on three Qwen-2B settings. Raw TopK and PCA at $k{=}160$ (PCA dim $=2k=320$); Enc-only and PUMA at $k{=}160$ except F200K at $k{=}128$. \textbf{Bold} = best per (data, metric); \underline{underline} = second best.}
\label{tab:baselines-2b}
\centering
\small
\setlength{\tabcolsep}{5pt}
\begin{tabular}{@{}llccc@{}}
\toprule
Data & Method & Hit@5 & R@10 & nDCG@10 \\
\midrule
\multirow{5}{*}{CIRR}        & Dense    & \underline{.5221} & \underline{.6281} & \underline{.4235} \\
                             & Raw TopK & .4655 & .5688 & .3769 \\
                             & PCA      & .5094 & .6124 & .4153 \\
                             & Enc-only & .4518 & .5614 & .3631 \\
                             & PUMA     & \textbf{.5329} & \textbf{.6449} & \textbf{.4356} \\
\midrule
\multirow{5}{*}{FashionIQ}   & Dense    & \underline{.2029} & \underline{.2680} & \textbf{.1706} \\
                             & Raw TopK & .1621 & .2168 & .1355 \\
                             & PCA      & .1891 & .2504 & .1574 \\
                             & Enc-only & .1667 & .2263 & .1367 \\
                             & PUMA     & \textbf{.2062} & \textbf{.2697} & \underline{.1681} \\
\midrule
\multirow{5}{*}{Fashion200K} & Dense    & .1198 & .0879 & .0792 \\
                             & Raw TopK & .0762 & .0621 & .0513 \\
                             & PCA      & .0861 & .0652 & .0538 \\
                             & Enc-only & \underline{.1495} & \underline{.1164} & \underline{.0967} \\
                             & PUMA     & \textbf{.1763} & \textbf{.1355} & \textbf{.1144} \\
\bottomrule
\end{tabular}
\end{table}

\cref{tab:trained-denseae-results} compares PUMA with a learned dense compressor at the same memory budget (TrAE). The trained dense autoencoder uses bottleneck dimension $2k$ and is trained on the same data with the same retrieval losses, optimization budget, and validation protocol as PUMA where applicable; sparsity-specific components such as AuxK, TopK annealing, and support-mask alignment are omitted or replaced by dense analogs. Across 15 model--dataset settings, the trained dense AE is below PUMA on 12. It wins only on MSCOCO for Qwen-2B/8B and on RZen--FashionIQ. Thus adaptation alone does not explain PUMA's gains: at the same per-vector budget, the sparse representation is usually the stronger retrieval representation.

\begin{table}[t]
\caption{Fixed-$k$ matched-memory trained dense AE across backbones. nDCG@10; \textbf{bold} = best of Dense/PUMA/TrAE, \underline{underline} = second best.}
\label{tab:trained-denseae-results}
\centering
\footnotesize
\setlength{\tabcolsep}{5pt}
\begin{tabular}{@{}llcccc@{}}
\toprule
Model & Data & $k$ & Dense & PUMA & TrAE \\
\midrule
\multirow{5}{*}{Qwen-2B}   & CIRR    & 160 & \underline{.4235} & \textbf{.4357} & .3885 \\
                           & FashIQ  & 160 & \textbf{.1706} & \underline{.1679} & .1327 \\
                           & VisN    & 160 & \textbf{.2256} & \underline{.2169} & .2125 \\
                           & F200K   & 128 & .0792 & \textbf{.1141} & \underline{.1058} \\
                           & MSCOCO  & 144 & .7181 & \underline{.7191} & \textbf{.7196} \\
\midrule
\multirow{5}{*}{Qwen-8B}   & CIRR    & 144 & \textbf{.4961} & \underline{.4835} & .4354 \\
                           & FashIQ  & 128 & \textbf{.1942} & \underline{.1922} & .1478 \\
                           & VisN    & 128 & .2296 & \textbf{.2561} & \underline{.2523} \\
                           & F200K   & 128 & .1048 & \textbf{.1322} & \underline{.1183} \\
                           & MSCOCO  & 128 & \underline{.7349} & .7297 & \textbf{.7367} \\
\midrule
\multirow{5}{*}{RZenEmbed} & CIRR    & 160 & \textbf{.4860} & \underline{.4712} & .4087 \\
                           & FashIQ  & 160 & \underline{.1952} & \textbf{.1994} & .1557 \\
                           & VisN    & 160 & \textbf{.3933} & \underline{.3588} & .3571 \\
                           & F200K   & 128 & .1339 & \textbf{.1578} & \underline{.1392} \\
                           & MSCOCO  & 144 & \textbf{.7630} & .7384 & \underline{.7445} \\
\bottomrule
\end{tabular}
\end{table}

Together, these controls rule out simpler explanations for PUMA's gains. Raw TopK shows that direct sparsification of dense coordinates is insufficient; matched-memory PCA shows that memory reduction alone is insufficient; the trained dense AE shows that supervised dense adaptation is usually insufficient; and encoder-only TopK shows that sparse retrieval supervision without a decoder is insufficient. The evidence instead supports the full recipe: sparse structure, decoder-backed reconstruction, and retrieval-aligned fine-tuning.

\subsection{Efficiency: Storage and Search Time}
\label{sec:results-efficiency}

PUMA reduces storage by representing each vector with only $k$ feature ids and weights. Under FP32 accounting, dense embeddings cost $4d$ bytes, while sparse codes cost $8k$ bytes with 32-bit ids and FP32 weights. At our operating points ($k\!\in\![128,160]$), this gives an $8\times$ reduction for Qwen-2B ($d{=}2048$) and up to $16\times$ for Qwen-8B ($d{=}4096$).

The $8k$ accounting is conservative. In a fixed index, feature ids only need to address atoms that appear in the candidate pool. Remapping corpus-active atoms at index construction leaves sparse vectors and retrieval scores unchanged, but reduces id width. In our diagnostics, at most 497 atoms are active, so 9-bit ids suffice; at $k{=}128$, this reduces PUMA storage from $1{,}024$ to $656$ bytes per vector. This remapping is optional, so the trend remains the deployment-agnostic $8$--$16\times$ reduction.

We also compare against quantized dense storage. FP16 and INT8 dense retrieval are essentially lossless in our measurements: on representative settings spanning CIRR, Fashion200K, and VisualNews at Qwen-2B and Qwen-8B, nDCG@10 changes by at most $0.001$ between FP32, FP16, and INT8 dense retrieval. For example, CIRR-2B gives $0.4239/0.4241/0.4236$ for FP32/FP16/INT8, and VisualNews-8B gives $0.2292/0.2292/0.2292$. Thus FP16 gives a free $2\times$ dense-storage reduction and INT8 gives a free $4\times$ reduction. PUMA remains smaller even against INT8 dense: with 9-bit ids and FP32 sparse weights, PUMA uses $656$ bytes per vector, compared with $2{,}048$ bytes for INT8 dense at Qwen-2B and $4{,}096$ bytes at Qwen-8B. Quantization is complementary to PUMA, since the sparse weights can also be quantized; thus, the FP32-weight accounting in \cref{tab:storage} is conservative.

\begin{table}[t]
\caption{
Per-vector storage in bytes at $k{=}128$. Compact ids remap logical atoms to corpus-active ids of width $\lceil\log_2 D_{\mathrm{active}}\rceil$ bits. The 9-bit row is a worst-case bound across our measured settings; the 8-bit row applies when $D_{\mathrm{active}}\le256$ (e.g., Qwen-8B CIRR).}
\label{tab:storage}
\centering
\small
\setlength{\tabcolsep}{2pt}
\begin{tabular}{@{}lrrr@{}}
\toprule
Representation & 2B (B) & 8B (B) & vs full dense \\
\midrule
Full dense ($4d$, FP32)       & 8192 & 16384 & $1\!\times$ \\
Full dense ($2d$, FP16)       & 4096 & 8192  & $2\!\times$ \\
Full dense ($d$, INT8)        & 2048 & 4096  & $4\!\times$ \\
Trained dense AE ($2k$, FP32) & 1024 & 1024  & 8 / 16$\,\times$ \\
\midrule
PUMA, 32-bit ids ($8k$)       & 1024 & 1024  & 8 / 16$\,\times$ \\
PUMA, 16-bit ids ($6k$)       & 768  & 768   & 10.7 / 21.3$\,\times$ \\
\textbf{PUMA, 9-bit ids}      & \textbf{656} & \textbf{656} & \textbf{12.5 / 25.0$\,\times$} \\
PUMA, 8-bit ids               & ---  & 640   & --- / 25.6$\,\times$ \\
\bottomrule
\end{tabular}
\end{table}

PUMA also improves exact search time on large candidate pools. On VisualNews 8B (542K candidates), exact dense search takes $423.8$\,s versus $17.3$\,s for sparse search ($24.5\times$ faster). On Fashion200K 8B (202K candidates), the gap is $13.8$\,s versus $0.77$\,s ($17.9\times$ faster). Dense search remains faster only on the much smaller MSCOCO pool, where sparse-kernel overhead dominates the low candidate count. These timings compare exact dense and exact sparse scoring to isolate the representation-level effect.

\subsection{Ablation: Which Components Matter?}
\label{sec:results-ablation}

We use leave-one-out ablations on two Qwen-2B tasks: CIRR for composed retrieval and Fashion200K for text-to-image retrieval. Each run removes one loss or stage while keeping the remaining training protocol fixed. The ablation uses a fixed $k{=}128$ to compare variants under a common support; main results use the validation-selected operating points in \cref{tab:main-results}.

\paragraph{Stage \ding{173} leave-one-out.}
We remove one Stage~\ding{173} loss at a time while keeping the rest of PUMA fixed (\cref{tab:loss-loo}). All three terms are load-bearing, but their importance differs by task. On CIRR, every single ablation drops PUMA below dense, with AuxK feature revival causing the largest loss. On Fashion200K, the Stage~\ding{173} contrastive blend is most important, while cross-modal alignment and AuxK still provide clear gains. Thus the full recipe is needed for the composed-retrieval win, whereas Fashion200K remains above dense.

\begin{table}[t]
\caption{Leave-one-out ablation at Qwen-2B. Stage~\ding{173} rows zero one term of \cref{eq:stage2}; Stage~\ding{174} OFF skips contrastive fine-tuning. Other settings match full PUMA. n@10 is nDCG@10; $\Delta$ is relative to PUMA.}\label{tab:loss-loo}
\centering
\small
\setlength{\tabcolsep}{7pt}
\begin{tabular}{@{}l c r@{\;\;}c r@{\;\;}c@{}}
\toprule
& & \multicolumn{2}{c}{\textbf{CIRR}} & \multicolumn{2}{c}{\textbf{Fashion200K}} \\
\cmidrule(lr){3-4}\cmidrule(lr){5-6}
Variant & $k$ & n@10 & $\Delta$ & n@10 & $\Delta$ \\
\midrule
Dense                & --  & .4235 &              & .0792 &              \\
\textbf{PUMA (full)} & 128 & \textbf{.4333} & ---       & \textbf{.1144} & ---       \\
\midrule
\multicolumn{6}{@{}l}{\emph{Stage~\ding{173} (\cref{eq:stage2})}} \\
$-\mathcal{L}_{\mathrm{align}}$ & 128 & .4072 & $-.026$ & .1009 & $-.014$ \\
$-\mathcal{L}_{\mathrm{aux}}$   & 128 & .3975 & $-.036$ & .0995 & $-.015$ \\
$-\mathcal{L}_{\mathrm{ctr}}^{(2)}$ & 128 & .4061 & $-.027$ & .0857 & $-.029$ \\
\midrule
\multicolumn{6}{@{}l}{\emph{Stage~\ding{174}:}} \\
Stage~\ding{174} OFF & 128 & .4140 & $-.019$ & .1050 & $-.009$ \\
\bottomrule
\end{tabular}
\end{table}

\paragraph{Stage \ding{174} leave-one-out.}
The Stage~\ding{174} row evaluates the Stage~\ding{173} checkpoint directly, without contrastive fine-tuning. On Qwen-2B, removing Stage~\ding{174} hurts both tasks: $-0.019$ nDCG@10 on CIRR and $-0.009$ on Fashion200K. At 8B, the effect depends on support availability. Removing Stage~\ding{174} slightly hurts FashionIQ ($0.1922\!\to\!0.1883$), but improves CIRR in the support-starved regime identified in \cref{sec:results-generalization}. This suggests that Stage~\ding{174} is useful when enough support is available, while the 8B CIRR reversal is another symptom of the support-unavailable failure mode.

\section{Conclusion}
\label{sec:conclusion}

We presented \textbf{PUMA}, a post-hoc recipe that converts frozen universal multimodal embeddings into compact sparse retrieval codes. On Qwen3-VL-Embedding-2B, PUMA is competitive with dense retrieval across five M-BEIR tasks, with significant gains on CIRR and Fashion200K, near parity on FashionIQ and MSCOCO, and a remaining gap on VisualNews. The gains are not explained
by dimensionality reduction alone: raw TopK, PCA, trained matched-memory dense autoencoders, and encoder-only sparse training all fall short in representative comparisons. The remaining challenge is to jointly ensure sufficient pre-TopK support and retrieval alignment of the active support, the two axes isolated by our failure diagnostics.

\clearpage
\section*{Limitations}
This work opens several directions for future study. First, our main experiments focus on the Qwen embedding family, with RZenEmbed included as a cross-family check. The transfer results show that PUMA is not Qwen-specific, but they also suggest that sparsification behavior depends on the geometry of the frozen backbone. Evaluating additional universal multimodal embedders would clarify how broadly the recipe applies.

Second, our failure analysis diagnoses two bottlenecks but does not solve them. In particular, the Qwen-8B CIRR setting shows that stronger dense backbones can become support-limited before TopK selection, while RZenEmbed VisualNews shows that available support is not always retrieval-aligned. Future work should develop objectives that jointly encourage sufficient dictionary coverage and retrieval-aligned active features.

Finally, PUMA is a post-hoc method and therefore inherits the representational properties of the frozen backbone. This is desirable for efficiency because the backbone is not retrained, but it also means that any domain imbalance or bias in the dense embedding space may propagate to the sparse codes. Deployment-sensitive settings should therefore audit the sparsified retriever alongside the original dense model.

\section*{Acknowledgments}
This work has been carried out while \textit{Matteo Attimonelli} was enrolled in the Italian National Doctorate on Artificial Intelligence run by Sapienza University of Rome in collaboration with \textit{Politecnico Di Bari}.
We acknowledge the CINECA award under the ISCRA initiative for the availability of high-performance computing resources and support. We acknowledge ISCRA for awarding this project access to the LEONARDO supercomputer, owned by the EuroHPC Joint Undertaking, hosted by CINECA (Italy). This publication was funded by the following projects: "New Kocs Era" (project code CDP001540; CUP B85H24002010007); "AURA – Augmented Unified Resource for Agents" (project code CDP001572);
This research was conducted as part of the ENRESFOOD project, funded under the FutureFoodS partnership. The work received funding from the Austrian Science Fund (FWF; 10.55776/KIN4661525), the Dutch Ministry of Agriculture, Fisheries, Food Security and Nature (BO-43-219-026), the Italian Ministry of University and Research (MUR), and the German Federal Ministry of Research, Technology and Space (031B1717).

\bibliography{references}

\clearpage
\appendix

\section{Bootstrap Confidence Intervals}
\label{app:bootstrap}

\begin{table}[h]
\caption{Representative 95\% bootstrap intervals for nDCG@10 on Qwen-2B. Rows show a PUMA win, near parity, and a dense win.}
\label{tab:bootstrap-selected}
\centering
\footnotesize
\setlength{\tabcolsep}{3pt}
\begin{tabular}{@{}lcc@{}}
\toprule
Data & Dense & PUMA \\
\midrule
CIRR   & .4233 [.4105,.4349] & \textbf{.4331 [.4214,.4446]} \\
FashIQ & .1705 [.1624,.1785] & .1681 [.1602,.1760] \\
VisN   & \textbf{.2258 [.2209,.2308]} & .2169 [.2122,.2218] \\
\bottomrule
\end{tabular}
\end{table}

\section{Support-Usage Diagnostics}
\label{app:support}
\cref{tab:support-diagnostics} reports pre-activation statistics for our checkpoints. Encoder-only training and the final SAE are both well past support saturation on 2B (mean positive pre-activations $\gg k$, exact realization of the requested $k$ on every example). On 8B, the SAE's positive pre-activations drop sharply, explaining the support-limited regime described in \cref{sec:results-generalization}.

\begin{table}[h]
\caption{Support-usage diagnostics ($k{=}128$). Pos-pre: mean positive pre-activations before TopK. Active: mean surviving features at native support. Reach: fraction of examples realizing $k$ exactly.}
\label{tab:support-diagnostics}
\centering
\small
\setlength{\tabcolsep}{4pt}
\begin{tabular}{@{}llccc@{}}
\toprule
Data & Setting & Pos-pre & Active & Reach \\
\midrule
CIRR & Enc-only 2B & 3774.4 & 128.0 & 1.000 \\
CIRR & PUMA 2B     & 363.8  & 128.0 & 1.000 \\
\midrule
CIRR & PUMA 8B     & 167.5  & 128.0 & 1.000 \\
\bottomrule
\end{tabular}
\end{table}

\section{Additional Ablation Study}
\begin{figure*}[t]
\centering
\includegraphics[width=\linewidth]{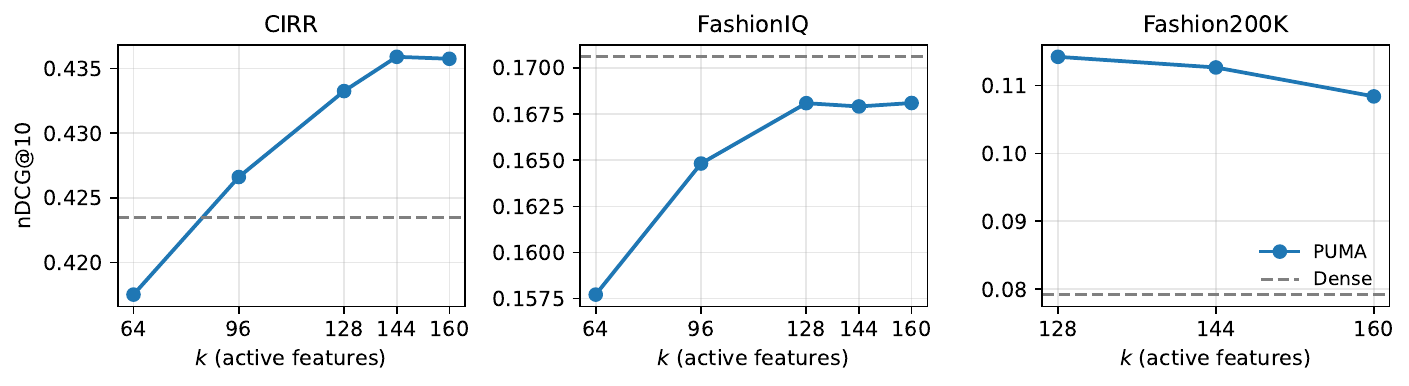}
\caption{PUMA nDCG@10 at Qwen-2B as the target sparsity $k$ varies, on CIRR, FashionIQ, and Fashion200K. Dashed line: dense baseline.}
\label{fig:k-sweep}
\end{figure*}

\paragraph{Effect of the target sparsity $k$.}
\Cref{fig:k-sweep} sweeps the inference-time support $k\!\in\!\{64,96,128,144,160\}$ at Qwen-2B on CIRR, FashionIQ, and Fashion200K. Quality saturates around $k\!=\!128$: increasing $k$ further yields diminishing returns, while $k\!=\!64$ already reaches near-saturation on FashionIQ and Fashion200K. CIRR is the most $k$-sensitive of the three (the most stringent regime, as expected for composed queries), but its curve also flattens past $k\!=\!128$. Combined with the storage analysis in \cref{sec:results-efficiency}, this places the practical operating point in the $k\!\in\![96,160]$ band: smaller $k$ gives a tighter index, larger $k$ recovers a tail of the dense ranking. Expansion ratio $D=16d$ is held fixed across all experiments; varying $D$ is left for future work.

\paragraph{Cross-modal alignment impact.}
The Stage~\ding{173} LOO above shows that removing $\mathcal{L}_{\mathrm{align}}$ costs retrieval quality, but does not show that the loss actually drives paired queries and candidates onto a common subset of dictionary atoms. To check this directly, we measure the Jaccard overlap of the TopK supports of paired and random query--candidate pairs on the trained PUMA-2B checkpoints (\cref{tab:jaccard-align}). On all three datasets the paired support overlap is substantially larger than the random baseline: $+0.177$ on CIRR, $+0.134$ on FashionIQ, and $+0.088$ on Fashion200K. The lift is largest on the two composed-retrieval datasets, which is where the alignment loss is supervised on genuinely cross-modal $(\mathbf{q},\mathbf{d}^+)$ pairs. Across all three datasets fewer than $500$ atoms ever fire, mirroring the concentration observed in \cref{tab:failure-diagnostics}, so the alignment behavior holds within an active dictionary that is two orders of magnitude smaller than the trainable $D=16d$.

\begin{table}[h]
\caption{Cross-modal TopK-support Jaccard on Qwen-2B PUMA ($k{=}128$, $D{=}32{,}768$). Paired: $(\text{query},\text{positive candidate})$ pairs. Random: query paired with a uniformly sampled candidate. Lift is paired $-$ random.}
\label{tab:jaccard-align}
\centering
\small
\setlength{\tabcolsep}{2.3pt}
\begin{tabular}{@{}lccccc@{}}
\toprule
Dataset & Pairs & Paired $J$ & Random $J$ & Lift & $|D_{\text{active}}|$ \\
\midrule
CIRR        & 4170 & 0.351 & 0.174 & $+0.177$ & 481 \\
FashionIQ   & 6003 & 0.314 & 0.180 & $+0.134$ & 424 \\
Fashion200K & 1719 & 0.211 & 0.123 & $+0.088$ & 447 \\
\bottomrule
\end{tabular}
\end{table}

\paragraph{Ultra-sparse regime ($k\!\le\!64$).}
Although our headline operating points sit in $k\!\in\![64,160]$, two questions remain: (i)~how steeply does PUMA degrade as $k$ shrinks toward the ultra-sparse regime used by SPLADE-style text retrievers, and (ii)~how cheap can the index be made before retrieval collapses? \cref{tab:small-k-sweep} extends the sweep with $k\!\in\!\{16,32,48,64\}$ on the same three 2B datasets. The curve is smooth: nDCG@10 at $k\!=\!16$ already captures $74\%$ of the dense score on CIRR ($0.313/0.4235$) and $69\%$ on Fashion200K ($0.054/0.0792$). By $k\!=\!48$ PUMA reaches $96\%$ of dense on CIRR ($0.407$) and \emph{exceeds} dense on Fashion200K ($0.110$ vs $0.079$, where dense is weak). The implication is a per-corpus storage knob: with $9$-bit ids, $k\!=\!16$ corresponds to $\sim\!82$ bytes per vector ($\sim\!100\times$ smaller than FP32 dense at 2B), trading $\sim\!25\%$ of dense quality on the hardest of the three tasks; $k\!=\!48$ at $\sim\!246$ bytes ($\sim\!33\times$) preserves it within $4\%$. The ultra-sparse rows therefore extend the sparsity-vs-quality frontier of \cref{fig:k-sweep} rather than break it.

\begin{table}[h]
\caption{Ultra-sparse PUMA at Qwen-2B (test split). Rows extend \cref{fig:k-sweep} below the $k\!=\!64$ point. All numbers measured from the same checkpoints used in \cref{tab:main-results}, with $k$ varied only at inference.}
\label{tab:small-k-sweep}
\centering
\small
\setlength{\tabcolsep}{5pt}
\begin{tabular}{@{}lcccc@{}}
\toprule
Setting & $k$ & Hit@5 & R@10 & nDCG@10 \\
\midrule
\multirow{5}{*}{CIRR}        & dense & .5221 & .6281 & .4235 \\
                              & 16 & .3784 & .4874 & .3133 \\
                              & 32 & .4540 & .5622 & .3743 \\
                              & 48 & .4887 & .6038 & .4070 \\
                              & 64 & .4990 & .6185 & .4177 \\
\midrule
\multirow{5}{*}{FashionIQ}   & dense & .2029 & .2680 & .1706 \\
                              & 16 & .1129 & .1566 & .0914 \\
                              & 32 & .1609 & .2152 & .1297 \\
                              & 48 & .1821 & .2401 & .1478 \\
                              & 64 & .1922 & .2580 & .1580 \\
\midrule
\multirow{5}{*}{Fashion200K} & dense & .1198 & .0879 & .0792 \\
                              & 16 & .0861 & .0799 & .0543 \\
                              & 32 & .1449 & .1326 & .0900 \\
                              & 48 & .1629 & .1556 & .1099 \\
                              & 64 & .1763 & .1654 & .1151 \\
\bottomrule
\end{tabular}
\end{table}

\section{Training Details}
\label{app:training-details}
We train SAEs for two stages on cached query/candidate embeddings from each M-BEIR task. Stage~\ding{173} (sparse pretraining) optimizes the objective in \cref{eq:stage2} with cosine reconstruction, retrieval-similarity supervision, AuxK feature revival, and a small Stage~\ding{173} contrastive blend; Stage~\ding{174} (retrieval fine-tuning) emphasizes InfoNCE on sparse codes with reconstruction and distillation kept as small regularizers. Both stages use AdamW with linear warmup and cosine decay, decoder column-norm renormalization after each step, and progressive $k$-annealing toward the target sparsity. For Qwen3-VL-Embedding-2B we use $D{=}32{,}768$; for 8B we use the analogously expanded $D{=}16d$. Checkpoints are selected on held-out sparse nDCG@10 evaluated on the validation split.

\paragraph{Loss weights.}
\cref{tab:loss-weights} lists all loss weights and their sources. The values are held \emph{fixed} across datasets and backbones — no per-task tuning. Each is motivated by a deliberate role rather than a grid search: $\lambda_a$ comes from MGSAE-style alignment defaults \citep{mgsae} and is moderate so the alignment/distillation signal shapes the support pattern without dominating reconstruction; $\lambda_x{=}1/32$ is the standard AuxK weight reported in TopK SAE work \citep{gao_topksae} and adopted by CSRv2 \citep{csr_v2}; $\lambda_c, \alpha, \beta$ are deliberately small ($\le 0.1$) so the auxiliary terms remain present during contrastive training without overwhelming the ranking objective; and $\tau{=}0.05$ is the common temperature in contrastive retrieval. In the retrieval-cache training mode used throughout the paper, $\mathcal{L}_{\mathrm{align}}$ and $\mathcal{L}_{\mathrm{dist}}$ collapse to a single retrieval-similarity distillation term, so a shared scalar $\lambda_a{=}\lambda_d$ controls both. The leave-one-out study in \cref{tab:loss-loo} validates this choice: each weight is load-bearing — zeroing any single one produces a measurable, smooth degradation rather than a sharp peak around the chosen value — which is the signature of moderate, untuned regularizers rather than over-fit hyperparameters.

\begin{table}[h]
\caption{PUMA loss weights, held fixed across all datasets and backbones.}
\label{tab:loss-weights}
\centering
\small
\setlength{\tabcolsep}{4pt}
\begin{tabular}{@{}lll@{}}
\toprule
Weight & Value & Source / role \\
\midrule
$\lambda_a{=}\lambda_d$ & $0.3$ & Alignment / dot-product distillation \\
$\lambda_x$ & $1/32$ & AuxK default \citep{gao_topksae} \\
$\lambda_c$ & $0.05$ & Stage~\ding{173} InfoNCE blend (small regularizer) \\
$\alpha$ & $0.1$ & Stage~\ding{174} reconstruction regularizer \\
$\beta$ & $0.05$ & Stage~\ding{174} distillation regularizer \\
$\tau$ & $0.05$ & InfoNCE temperature \\
\bottomrule
\end{tabular}
\end{table}

\paragraph{Optimization.}
AdamW with weight decay $10^{-2}$. Learning rates: $3{\times}10^{-4}$ (Stage~\ding{173}), $1{\times}10^{-4}$ (Stage~\ding{174}), each with $1{,}000$ / $500$ linear warmup steps followed by cosine decay. Stage~\ding{173} runs $40{,}000$ steps; Stage~\ding{174} runs $5{,}000$ steps. Batch size $64$ throughout.
Validation-based checkpoint selection uses a $10\%$ slice of the training pairs evaluated on sparse nDCG@10 at the target $k$.

\end{document}